# Feasibility Study of Real-Time Planning

# for Stereotactic Radiosurgery

[1]Qinghui Zhang, Ph.D., [1]Yulin Song, Ph. D., [1]Maria Chan, Ph.D., [1]Chandra Burman, Ph.D.,

[2]Yoshiya Yamada, MD.

[1]*Department of Medical Physics* and [2]*Department of Radiation Oncology*

*Memorial Sloan-Kettering Cancer Center, New York, NY 10065*

Corresponding author:

Qinghui Zhang, Ph.D.

Department of Medical Physics

Memorial Sloan-Kettering Cancer Center

1275 York Ave

New York, NY 10065

Phone: (212) 639-6036

**Email: qinghui.zhang@gmail.com**

**Acknowledgements:** We would like to thank Gig Mageras and Margie Hunt for helpful discussions on this paper. We also thank the referee who brought Reference [15] to our attention.

**ABSTRACT**

**Purpose:** 3D rotational setup errors in radiotherapy are often ignored by most clinics due to inability to correct or simulate them accurately and efficiently. There are two types of rotation-related problems in a clinical setting. One is to assess the affected dose distribution in real-time if correction is not applied and the other one is to correct the rotational setup errors prior to the initiation of the treatment. Here, we present our analytical solutions to both problems.

**Methods**: 1) To assess the real-time dose distribution, eight stereotactic radiosurgery (SRS) cases were used as examples. For each plan, two new sets of beams with different table, gantry, and collimator angles were given in analytical forms as a function of patient rotational errors. The new beams simulate the rotational effects of the patient during the treatment setup. By using one arbitrary set of beams, SRS plans were re-computed with a series of different combinations of patient rotational errors, ranging from $(-5^0, -5^0, -5^0)$ to $(5^0, 5^0, 5^0)$ (roll, pitch, and yaw) with an increment of $1^0$ and compared with those without rotational errors. For each set of rotational errors, its corresponding equivalent beams were computed using our analytical solutions and then used for dose calculation. 2) To correct for the rotation rotational errors, two new sets of table, gantry, and collimator angles were derived analytically to validate the previously-published derivation. However, in our derivation, a novel methodology was developed and two sets of table, gantry, and collimator angles were obtained in analytical forms. The solutions provide an alternative approach to rotational error correction by rotating the couch, gantry, and collimator rather than the patient.

**Results:** For demonstration purpose, the above-derived new beams were implemented in a treatment planning system (TPS) to study the rotational effects on the SRS cases. For each case, we have generated 10 additional plans that accounted for different rotations of the patient. We have found that rotations have an insignificant effect on the minimal, maximum, mean doses, and $V_{80\%}$ of the planning target volume (PTV) when the rotations were relatively small. This was particularly true for the small and near-spherical targets. They, however, did change $V_{95\%}$ significantly when the rotations approached $5^0$. Our theory has been validated with clinical SRS cases and proven to be practical and viable. Our preliminary results demonstrate that the rotational effects are patient-specific and depend on several important factors, such as the PTV size, the PTV location, and the beam configuration. The solutions given in this paper are of great potential values in clinical applications.

**Conclusions:** We have derived the analytical solutions to a new set of table, gantry, and collimator angles for a given treatment beam configuration as a function of patient rotational errors. One solution was used to assess the dosimetric effects of an imperfect patient setup and the other one was used to correct for the setup errors without rotating the patient. Compared to the widely-adopted method of rotation effect assessment by importing the rotational CT images into TPS, our equivalent beam approach is simple and accurate. The analytical solutions to correcting for rotational setup errors prior to treatment were also derived. Based on our initial clinical investigations, we firmly believe that clinically viable real-time treatment planning and adaptive radiation therapy are feasible with this novel method.

## I. INTRODUCTION

It is well known that patient setup errors prior to treatment compromise the target localization precision and, thus, the quality of radiation therapy. To improve the accuracy of patient positioning, the dedicated on-board cone beam CT imaging systems (CBCT) [1], ExacTrac [2], and AlignRT [3-5] have been developed and widely used in modern cancer centers. Though, mathematically, six degrees-of-freedom (DOF) registration of CBCT and planning CT can yield accurate translational and rotational shifts by using a rigid-body transformation model [6], in reality, the majority of the mainstream LINAC couches are not equipped with the necessary hardware to perform rotational corrections either automatically or manually except for the horizontal rotation. Recently, several leading vendors have developed six-DOF couches [7], which, however, are not currently available at many cancer centers due to unjustifiable high capital cost. Consequently, the residual setup errors have to be incorporated into the clinical target volume (CTV) when designing a clinically suitable planning target volume (PTV) [8-12]. This extra margin, composed of normal healthy tissue, will always absorb additional amount of unnecessary radiation and could lead to an increased risk of complications. In addition, even with well-designed six-DOF couches, it is widely recognized that the residual rotational errors still exist due to a variety of reasons. Therefore, the accurate estimation, precise correction, and, in particular, the instant evaluation of the dosimetric impact of the residual setup errors remain clinically relevant and constitute a challenging mathematical problem in radiation treatment planning.

Based on the endpoints, the solutions to the rotational errors can be classified into two categories. One is to assess their dosimetric effects either through an effective real-time planning technique or a post-treatment off-line re-planning method for subsequent adaptive treatment. The other one is to seek a clinically viable procedure to correct them with negligible residual errors. For the dosimetric effect assessment, the common approach is to simulate the patient rotation by rotating the CT images in the TPS. This operation is both inefficient and error-prone without special software. Even with the aid of the special software, the rotation operation will inevitably result in non-negligible numerical error because of the non-continuous nature of the contours and the images. Beside this, extra buffer and memory space are required to store the rotated images and contours. For multi-fraction treatments, multiple CT image and contour rotations would have to be performed and could become a very computationally intensive task, which would significantly increase the demand on the memory space and computing speed.

For adaptive dosimetry calculations, the planner also needs to compute the composite dosimetry of all previously-treated fractions. This would require prior rotation of the planning CT images and contours in the TPS, rendering the rotation method unsuitable for real-time planning in a busy clinical environment. Even worse, when summing doses from all treated fractions, each of them would have to be rotated back to its original planning CT position, thus doubling the computation time and error. An alternative strategy to estimate the rotation-induced dosimetric effect could be the real-time planning based on the setup CBCT. However, this technology is still in its cradle phase and is not fully implementable at this point due to various reasons. Among them, the most pronounced one is that the CT number of the CBCT is not precisely defined and calibrated as compared to the diagnostic CT scanners. Consequently, the tissue relative electron densities derived from the CBCT CT numbers are inaccurate for dose calculation. An additional bottleneck associated with CBCT-based planning is the extensive overhead work and long pre-optimization processing time, which makes this technique unsuitable as a viable real-time planning modality in a clinical environment. Furthermore, the scan range of CBCT is very limited and, thus, not sufficient for most of the non-SRS planning.

For setup error correction (the second problem), one of the existing techniques is to use a set of different couch, gantry, and collimator angles computed from the formulas given by Yue et al [13]. However, Yue et al only obtained a single solution to the underlying mathematical problem and their solution cannot be used for dosimetric effect estimation.

In this paper, we provide an alternative approach to assess the rotation-induced dosimetric effects and real-time planning by using the original planning CT. This is accomplished by adjusting the gantry, collimator, and couch angles of the original treatment beams to simulate the patient rotation so that the beams' eye views (BEVs) of these new beams are the same as those of the original beams when the patient has certain rotational errors. This new set of the gantry, collimator, and couch angles is computed by using the novel analytical solutions to be described in this paper.

In addition, the analytical expressions of the gantry, collimator, and couch angles used to correct for the rotational setup errors have also been derived by using a different method. Mathematically, our results not only validate the formulas given previously by Yue et al, but also provide two equivalent solutions. Furthermore, our derivation adopts the notations developed by Siddon [14]. Thus, it is easy to follow and understand.

Clinically, our technique is able to reduce the planning time to an acceptable level so that real-time planning becomes feasible. Using our method, the newly-computed dosimetry can be used as a new baseline for subsequent adaptive dosimetry calculations for multi-fraction cases. It can also be used to estimate the possible dosimetric errors caused by the setup rotations for single-fraction cases. Dosimetrically, our approach is accurate and straightforward. In particular, any potential errors made during the rotation of the CT images and contours can be avoided and all our calculations are performed on the planning CT images only even for cases required multiple treatments. Operationally, the physicist can use the setup rotational errors measured during the patient setup to calculate a new equivalent beam for each original treatment beam using the method provided in this paper, which will then be used to calculate an updated patient dosimetry online.

In summary, the complete analytical solutions (two) to the specific two problems introduced above have been derived. More significantly, our solutions can be directly used for rotation-induced dosimetric error estimation and real-time planning. In this paper, we present the detailed derivations of our mathematical theories and, as a demonstration, the preliminary results of their applications on clinical SRS cases.

## II. METHODS

According to Siddon's definition [14], the lab coordinate system (fixed system) is an orthogonal, right-handed coordinate system that is fixed in the treatment room. The origin is at the isocenter. The $X_{Lab}$ axis is towards right when facing the gantry and the $Y_{Lab}$ axis is the direction towards the gantry. The $Z_{Lab}$ is towards the treatment room ceiling. The gantry, collimator, and couch coordinate systems are fixed to the gantry, the collimator, and the couch, respectively. They are defined in such a way that they are in coincidence with the lab coordinate system when gantry angle, collimator angle, and couch angle are zeros. The gantry rotates about the $Y_{Lab}$-axis according to the right-hand-rule. Similarly, the collimator rotates about the $Z_{Gantry}$ axis according to the right-hand-rule. For the couch, following Siddon's convention, it rotates about the $Z_{Lab}$ according to the left-hand-rule. By using these definitions, Siddon has derived the following relationship

$$\vec{V}_C = R_Z(\theta_C) R_Y(\theta_G) R_Z(\theta_T) \vec{V}_T \qquad (1)$$

Here, $\vec{V}_T$ and $\vec{V}_C$ are patient coordinates in the couch and collimator systems, respectively. In other words, $\vec{V}_C$ is a beam's eye view of the patient coordinate that has the value of $\vec{V}_T$ in the couch coordinate system. Here, $R_Z(\theta), R_Y(\theta)$, and $R_X(\theta)$ are the unitary rotation matrices that are defined in the appendix of this paper (which can also be found in Siddon's paper).

It is interesting to point out here that if

$$\theta_G \rightarrow 2\pi - \theta_G, \ \theta_C \rightarrow \pi + \theta_C, \theta_T \rightarrow \pi + \theta_T \qquad (2),$$

Eq. (1) remains unchanged. This means that we have two equivalent beam arrangements for a given plan.

Currently, CBCT is widely used in the patient setup. If a patient setup is perfect, then the patient position on the couch is the same as that in the planning CT. Assuming that the tumor isocenter is at the origin of the couch coordinate (isocenter of the linac), we define the CBCT and the planning CT coordinate systems are the same as that of the couch (but the CBCT and planning CT coordinates are defined to be fixed in the patient).

$$\vec{V}_T = \vec{V}_{CBCT} = \vec{V}_{CT} \tag{3}$$

However, patient's rotation always exists during the setup. Suppose that the rotation angles between the CBCT images and CT images are $\theta_x, \theta_y, \theta_z$, defined as the rotation angles from the CT image to the CBCT image and following the right-hand-rule. Then, the CBCT (the patient) in the couch system can be expressed as

$$\vec{V}_T = R_X(-\theta_x)R_Y(-\theta_y)R_Z(-\theta_z)\vec{V}_{CT} \tag{4}$$

Here, $\vec{V}_{CT}$ represents the anatomy point in the CT images. From beam's eye view equation (Eq. (1)), for a perfect patient setup, we have

$$\vec{V}_C = R_Z(\theta_C)R_Y(\theta_G)R_Z(\theta_T)\vec{V}_{CT} \tag{5}$$

When there are rotations, $\vec{V}_C$ can be computed by plugging Eq. (4) into Eq. (1):

$$\vec{V}_C = R_Z(\theta_C)R_Y(\theta_G)R_Z(\theta_T)R_X(-\theta_x)R_Y(-\theta_y)R_Z(-\theta_z)\vec{V}_{CT} \tag{6}$$

It is clear that due to rotations, the beam's eye view of the patient is different from that for the plan, as illustrated by Eq. (6) and Eq. (5), respectively.

There are two types of problems associated with Eq. (5) and Eq. (6). (1): a new set of beams in Eq. (5) can be found such that their BEVs are the same as those expressed by Eq. (6). Therefore, the dosimetric effects of rotation can be studied. (2): To correct for the rotational effects in the setup, a new set of beams is needed in Eq. (6) such that Eq. (6) and Eq. (5) are equal. This is called the beam re-arrangement for the rotation correction. Both problems will be solved in this paper. In the derivation, we will use explicit matrix format. For clarity and simplicity, we put all those matrix expressions in the appendix of this paper.

**II.A** A new set of beams to study dosimetric effects

Real-time planning is not possible at present time because of long planning time. However, in this paper, we provide a practical method that can be potentially used for real-time planning. It can also be used for adaptive dosimetry correction for multi-fraction cases. Therefore, from a theoretical point of view, it would be valuable to be able to study the rotational effects using a conventional TPS. Nevertheless, the rotational effects cannot be studied by any of the commercial TPSs at present time. Almost all TPSs do not possess this functionality. As has been stated above, it is time consuming and error-prone to study the rotational effects by generating CT images with different rotation angles in the TPS. Here, we would like to propose a different method that can be easily implemented in the TPS. To do so, we derive a new set of beams such that their BEVs with respect to the planning CT are the same as the BEVs of the original beams for the patient with the same amount of rotational errors in the treatment room, In terms of BEVs, this new set of beams produces the same BEV effect as rotating the planning CT in the TPS or the patient on the treatment couch. In addition, BEVs are different for different beams, with each different original beam having its own new set of different gantry, couch, and collimator angles. Therefore, one can use this new set of beams to assess the rotational effects on dose distribution. From Eq. (5), we have

$$\vec{V}_C = R_Z(\theta_C^{new1}) R_Y(\theta_G^{new1}) R_Z(\theta_T^{new1}) \vec{V}_{CT} \qquad (7)$$

Eq. (7) has the same BEV as the original plan, but with the patient rotations incorporated. In other words, Eq. (7) is the same as Eq. (6) above.

Therefore,

$$\vec{V}_C = R_Z(\theta_C)R_Y(\theta_G)R_Z(\theta_T)R_X(-\theta_x)R_Y(-\theta_y)R_Z(-\theta_z)\vec{V}_{CT} = R_Z(\theta_C^{new1})R_Y(\theta_G^{new1})R_Z(\theta_T^{new1})\vec{V}_{CT} \quad (8)$$

Which can be further simplified as:

$$R_Z(\theta_C^{new1})R_Y(\theta_G^{new1})R_Z(\theta_T^{new1}) = R_Z(\theta_C)R_Y(\theta_G)R_Z(\theta_T)R_X(-\theta_x)R_Y(-\theta_y)R_Z(-\theta_z) \quad (9)$$

we denote

$$T^{new1} = R_Z(\theta_C^{new1})R_Y(\theta_G^{new1})R_Z(\theta_T^{new1}), \quad (10)$$

$$T = R_Z(\theta_C)R_Y(\theta_G)R_Z(\theta_T), \quad (10a)$$

and

$$R = R_X(-\theta_x)R_Y(-\theta_y)R_Z(-\theta_z) \quad (11)$$

Thus, Eq. (9) can be re-written as

$$T^{new1} = TR \quad (9a)$$

whose solutions are given in the appendix of this paper. This novel approach to estimate the rotational effects on dose distribution can be easily implemented in any TPSs. In the above, two new sets of equivalent beams have been derived. They can be implemented clinically to calculate an updated patient dose distribution even there are rotations in the setup.

**II.B** A new set of beams to correct for rotational effects

In this section, two new sets of equivalent beams will be derived. They can be used clinically to correct for the rotational effects such that the patient will receive the prescribed dose even with rotations present in the set up. However, this is only true under ideal conditions. In real clinical situations, beam profiles are machine hardware accuracy-dependent. Different from the previous method, which is a recalculation of the dose distribution in the TPS and is always doable, this method has to be applied in the treatment room prior to beam-on. Therefore, it is limited by the hardware constraints, such as clearance and potential collision. Thus, this method might not be implementable for some beam configurations. Nevertheless, we still provide solutions to this mathematical problem. We need to point out here that a solution has already been provided by Yue et al. [13]. But in this paper, we will derive two equivalent solutions and the method in this paper is different from that in Yue et al.'s paper.

To solve this problem, we need a new set of gantry, collimator and couch angles, $\theta_C^{New}, \theta_G^{New}$, and $\theta_T^{New}$ such that Eq.(6) is the same as Eq.(5), Thus, Eq. (6) changes to

$$\vec{V}_C = R_z(\theta_c^{new})R_Y(\theta_G^{new})R_Z(\theta_T^{new})R_X(-\theta_x)R_Y(-\theta_y)R_Z(-\theta_z)\vec{V}_{CT} \qquad (6A)$$

and we have

$$\vec{V}_C = R_z(\theta_c^{new})R_Y(\theta_G^{new})R_Z(\theta_T^{new})R_X(-\theta_x)R_Y(-\theta_y)R_Z(-\theta_z)\vec{V}_{CT} = R_z(\theta_C)R_Y(\theta_G)R_Z(\theta_T)\vec{V}_{CT} \qquad (12)$$

In other words, we try to find new $\theta_C^{new}, \theta_G^{new}, \theta_T^{new}$ such that

$$R_z(\theta_c^{new})R_Y(\theta_G^{new})R_Z(\theta_T^{new})R_X(-\theta_x)R_Y(-\theta_y)R_Z(-\theta_z) = R_z(\theta_C)R_Y(\theta_G)R_Z(\theta_T) \qquad (13)$$

or

$$R_Z(\theta_c^{new})R_Y(\theta_G^{new})R_Z(\theta_T^{new}) = R_z(\theta_C)R_Y(\theta_G)R_Z(\theta_T)R_Z(\theta_z)R_Y(\theta_y)R_X(\theta_x) \qquad (14)$$

We should point out that $R_X, R_Y, R_Z$ are unitary matrices. The multiplication of unitary matrices is still a unitary matrix. We denote

$$T = R_Z(\theta_C)R_Y(\theta_G)R_Z(\theta_T) \qquad (15)$$

and

$$T^{new} = R_Z(\theta_C^{new})R_Y(\theta_G^{new})R_Z(\theta_T^{new}) \qquad (16)$$

It is important to point out here that if $\theta_G^{new} \to 2\pi - \theta_G^{new}$, $\theta_C^{new} \to \pi + \theta_C^{new}$, $\theta_T^{new} \to \pi + \theta_T^{new}$, Eq. (16) remains unchanged. This means that we have two equivalent beam arrangements for a given plan.

Eq. (14) can be re-written as

$$T^{new} = TR^T \tag{14a}$$

The solutions to Eq. (14a) can be easily found (see Appendix).

By using this new set of beams, one can correct for the rotational errors and the patient will receive the prescribed dose as planned without physically repositioning the patient. Different from Yue et. al's method [13], our method is easy to follow and yields two sets of solutions.

## III. RESULTS AND DISCUSSION

Our primary interest in this study was to calculate the patient dose distribution with rotations because it is clinically implementable and applicable to all cases. To demonstrate our algorithms, we have evaluated eight SRS cases retrospectively. For each patient, we set rotation angles as $\theta_x = \theta_y = \theta_z$, ranging from $-5^0$ to $5^0$ with an angle increment of $1^0$. Then, we calculated the new set of gantry, collimator and couch angles for each rotation combination. The rotational errors presented in this paper represent the worst cases of scenario encountered in daily patient setups. Asymmetric rotation errors are more commonly observed in reality. However, we intentionally chose them to limit our data analysis to a manageable level. Otherwise, the combinations of rotation would be immense (125 in total) for dose calculations (1000 times for 8 cases). The data presented in the paper were the results of our initial investigations and were intended to be for demonstration purpose only. Our primary focus was on the derivations of the mathematical algorithms. Nevertheless, for asymmetric angle combinations, their dosimetric impact can be estimated in this way: for example, for a rotation angle combination of (2°, 3°, 4°), its induced dosimetric error would lie between those of (4°, 4°, 4°) and (2°, 2°, 2°). Our clinical SRS cases normally consist of 10 non-coplanar beams, thus, producing a set of 10 equivalent beams. As shown earlier, we have two equivalent solutions for each individual beam. We randomly selected one for our study. The selected equivalent beam may not be physically deliverable due to potential collision with the couch or the immobilization device. However, it is always possible for dose calculation in a TPS and, thus, will not affect the validity of this study. For each patient, we generated a total of 11 plans. For each plan, we have calculated the minimal dose, the mean dose, the maximal dose, the conformity index for the PTV (defined as the ratio of the volume receiving the prescribed dose to the PTV), $V_{80\%}$ (the percent PTV receiving 80% of the prescribed dose) and $V_{95\%}$ (the percent PTV receiving 95% of the prescribed dose). We have also calculated the minimal dose, the mean dose, the maximal dose, and $V_{95\%}$ for the CTV (the percent CTV receiving 95% of the prescribed dose). For our routine clinical treatments, the PTV dose is prescribed to the 80% isodose surface that covers entire PTV. We calculated all these dosimetric parameters for different rotation angles and their corresponding ratios with the original plan values (without rotations). The results are presented in Figs. (1-10). It is evident from the figures that rotational effects on the PTV and the CTV dose distributions are not dramatic for the SRS cases when the rotation angles are not so large. This is due to the near spherical shapes of the head and the PTV. Most importantly, the tumor size for these SRS cases is small, which makes the rotational effects on patient dose distribution much less pronounced. However, we found that the conformity index changed appreciably for patient 4, this is because the tumor in this case was located at the superficial left-posterior side of the head, far away from the central axis of the brain. Therefore, rotation produced a larger effect on the PTV than other cases. The tumor volume and position are listed in Table 1. In fact, the dosimetric effects of rotation depend on many factors, such as the beam number, beam arrangement in space, PTV size, PTV geometry, PTV location, plan quality, and so on. Therefore, it is difficult to determine the dominant one without performing an analysis of variance (ANOVA) for multiple factors, a task that is beyond the scope of this paper. Comparatively, $V_{95\%}$ was affected more significantly for all the patients, indicating a higher sensitivity to rotations. Because the CTV is always smaller than the PTV, it is not surprising that CTV $V_{95\%}$ is less affected by rotations than the PTV $V_{95\%}$. For the same reason, the maximum doses for the PTV and CTV showed very similar characteristics, as clearly demonstrated by Fig. 9. More importantly, we have found that the rotational effects are patient-specific. Therefore, a novel, but viable method like this is indispensible in this field to study the rotation -induced patient-specific dosimetric effects. It is interesting to point out that a different study by Yue et al [15] addressed the similar problem using a similar approach. The authors there have attempted to re-establish the previous treatment dose for a patient such that the dose would be included in the design of the new treatment for the same patient. This is another useful clinical application of this approach.

Practical Considerations for Clinical Implementations**:**

  To efficiently implement our method in real clinical situations, we suggest: 1) the physicist pre-calculates the dosimetry results for the typical range of setup rotational errors for a patient as part of the treatment planning process. This will create a patient-specific lookup table, i.e., dosimetry error vs. rotational errors and determine a suitable action threshold. The physicist and radiation oncologist will then decide whether a rotation correction is needed based on measured patient rotation angles. In the future, a population-based table could be constructed by combining and consolidating these patient-specific tables. 2) Although the two problems are separated mathematically, they could be combined together in clinical implementation. For example, if the treating physicist finds out that a correction is needed because of the large rotation angles. The treating therapist (under the supervision of a physicist) can perform the couch, gantry, and collimator angle corrections if there is no collision issue. Otherwise, the physicist needs to compute a real-time plan by using our first method in case of a potential collision.

## IV. CONCLUSIONS

We have derived analytical solutions of new couch, gantry and couch angles for a given beam configuration as a function of the patient rotation angles on the treatment couch. One of their potential applications is to assess the patient rotation-induced dosimetric error. These two analytical solutions can be easily implemented in any TPSs to determine the rotational effects on dose distribution. Our theory has been demonstrated with clinical SRS cases and proven to be practical and viable. For the SRS cases studied, we have found that the dose distribution has been perturbed for the PTV owing to rotations. However, they do not significantly affect the plan quality based on the currently prevailing planning standards. Since the implementation time for this method is very short, real-time planning in a clinical setting becomes possible with the proposed methodology. In our calculations, the original planning CT images were used, therefore, adaptive dosimetry calculations were straightforward and greatly simplified. The other potential application of our algorithms is to correct for the setup errors without physically rotating the patient. Our derivations are different from those in previous publications and are easy to follow. Two solutions to this mathematical problem were obtained. These novel setup rotation correction strategies are particularly relevant to cases where complex immobilization devices are used and that are treated on LINACS with a 4-DOF couch. To the best of our knowledge, none of the existing treatment planning systems contains a similar function, thus making this work particularly significant and valuable. The methodologies presented in this paper, of course, can be naturally extended to other disease sites, but those applications are out of the scope of our current investigation. For the demonstration purpose, SRS cases were studied retrospectively. However, because the primary objective of this paper was to derive the analytical solutions to the long existing mathematical problems, we did not pursue a comprehensive statistical study on a large sample size to determine the most rotation-sensitive factor. Nevertheless, clinical observations indicate that the rotation-induced dosimetric effects depend on many factors, including the tumor volume, tumor shape, body shape, number of beams, beam configuration in space, treatment modality, and the plan quality. In general, they are patient-specific. Undoubtedly, such a systematical statistical study of those factors on the patient dosimetry is clinically important and will be our future research projects.

**APPENDIX: The solutions of the new beam arrangements**

Rotations matrices are widely used in both physics and mathematics world. There are two kinds of rotation matrixes. One is the rotation of object relative to fixed axes and the other is rotation of the axes. In this paper, rotation of the axes is used. The coordinate system rotations of the x-, y-, and z-axes in a counterclockwise direction (right-hand rule) when looking towards the origin give the matrices

$$R_X(\theta) = \begin{bmatrix} 1 & 0 & 0 \\ 0 & \cos(\theta) & \sin(\theta) \\ 0 & -\sin(\theta) & \cos(\theta) \end{bmatrix}, R_Y(\theta) = \begin{bmatrix} \cos(\theta) & 0 & -\sin(\theta) \\ 0 & 1 & 0 \\ \sin(\theta) & 0 & \cos(\theta) \end{bmatrix}, R_Z(\theta) = \begin{bmatrix} \cos(\theta) & \sin(\theta) & 0 \\ -\sin(\theta) & \cos(\theta) & 0 \\ 0 & 0 & 1 \end{bmatrix} \quad (A.1)$$

Here $R_Z(\theta), R_Y(\theta)$ and $R_X(\theta)$ are unitary rotation matrix. If the object is rotated in a counter-clockwise direction but frame is fixed, by changing $\theta \to -\theta$, the rotations of object matrices are obtained. We denote

$$T = \begin{bmatrix} T_{11} & T_{12} & T_{13} \\ T_{21} & T_{22} & T_{23} \\ T_{31} & T_{32} & T_{33} \end{bmatrix} = R_Z(\theta_C) R_Y(\theta_G) R_Z(\theta_T)$$

$$= \begin{bmatrix} \cos\theta_C \cos\theta_G \cos\theta_T - \sin\theta_T \sin\theta_C & \cos\theta_C \cos\theta_G \sin\theta_T + \sin\theta_C \cos\theta_T & -\sin\theta_G \cos\theta_C \\ -\sin\theta_C \cos\theta_G \cos\theta_T - \cos\theta_C \sin\theta_T & -\sin\theta_C \cos\theta_G \sin\theta_T + \cos\theta_C \cos\theta_T & \sin\theta_G \sin\theta_C \\ \sin\theta_G \cos\theta_T & \sin\theta_G \sin\theta_T & \cos\theta_G \end{bmatrix}. (A.2)$$

For a new set of gantry angles, table angles, and couch angles $\theta_G^{new}, \theta_T^{new}, \theta_C^{new}$ and

$\theta_G^{new1}, \theta_T^{new1}, \theta_C^{new1}$, we can define a similar matrix as Eq. (A.2)

$$T^{new} = \begin{bmatrix} T_{11}^{new} & T_{12}^{new} & T_{13}^{new} \\ T_{21}^{new} & T_{21}^{new} & T_{23}^{new} \\ T_{31}^{new} & T_{32}^{new} & T_{33}^{new} \end{bmatrix} = R_Z(\theta_C^{new}) R_Y(\theta_G^{new}) R_Z(\theta_T^{new}) \qquad (A.3)$$

and

$$T^{new1} = \begin{bmatrix} T_{11}^{new1} & T_{12}^{new1} & T_{13}^{new1} \\ T_{21}^{new1} & T_{21}^{new1} & T_{23}^{new1} \\ T_{31}^{new1} & T_{32}^{new1} & T_{33}^{new1} \end{bmatrix} = R_Z(\theta_C^{new1}) R_Y(\theta_G^{new1}) R_Z(\theta_T^{new1}) \quad (A.4)$$

We define R matrix as

$$R = \begin{bmatrix} a_{11} & a_{12} & a_{13} \\ a_{21} & a_{22} & a_{23} \\ a_{31} & a_{32} & a_{33} \end{bmatrix} = R_X(-\theta_x) R_Y(-\theta_y) R_Z(-\theta_z)$$

$$= \begin{bmatrix} \cos\theta_y \cos\theta_z & -\cos\theta_y \sin\theta_z & \sin\theta_y \\ \sin\theta_x \sin\theta_y \cos\theta_z + \cos\theta_x \sin\theta_z & -\sin\theta_x \sin\theta_y \sin\theta_z + \cos\theta_x \cos\theta_z & -\sin\theta_x \cos\theta_y \\ -\cos\theta_x \sin\theta_y \cos\theta_z + \sin\theta_x \sin\theta_z & \cos\theta_x \sin\theta_y \sin\theta_z + \sin\theta_x \cos\theta_z & \cos\theta_x \cos\theta_y \end{bmatrix} \quad (A.5)$$

The solutions of Eq. (9) are:

For the case $\sin \theta_G^{new1} > 0$,

$$\theta_G^{new1} = \arccos[T_{31}a_{13} + T_{32}a_{23} + T_{33}a_{33}] \quad (A.6),$$

$$\theta_C^{new1} = \begin{cases} \arccos(-\dfrac{T_{11}a_{13} + T_{12}a_{23} + T_{13}a_{33}}{\sin \theta_G^{new1}}), & \text{if } T_{21}a_{13} + T_{22}a_{23} + T_{23}a_{33} > 0 \\ 2\pi - \arccos(-\dfrac{T_{11}a_{13} + T_{12}a_{23} + T_{13}a_{33}}{\sin \theta_G^{new1}}), & \text{if } T_{21}a_{13} + T_{22}a_{23} + T_{23}a_{33} < 0 \end{cases} \quad (A.7)$$

$$\theta_T^{new1} = \begin{cases} \arccos(\dfrac{T_{31}a_{11} + T_{32}a_{21} + T_{33}a_{31}}{\sin \theta_G^{new1}}), & \text{if } T_{31}a_{12} + T_{32}a_{22} + T_{33}a_{32} > 0 \\ 2\pi - \arccos(\dfrac{T_{31}a_{11} + T_{32}a_{21} + T_{33}a_{31}}{\sin \theta_G^{new1}}), & \text{if } T_{31}a_{12} + T_{32}a_{22} + T_{33}a_{32} < 0 \end{cases} \quad (A.8)$$

It has another set of solution, which is:

$$\theta_G^{new1} \to 2\pi - \theta_G^{new1}, \quad \theta_C^{new1} \to \theta_C^{new1} + \pi, \quad \theta_T^{new1} \to \theta_T^{new1} + \pi$$

For the special case when $\cos \theta_G^{new1} = \pm 1$, then $\theta_G^{new1} = 0, \pi$.

For $\theta_G^{new1} = 0$,

$$\theta_C^{new1} + \theta_T^{new1} = \begin{cases} \arccos(T_{21}a_{12} + T_{22}a_{22} + T_{23}a_{32}) & \text{if } T_{11}a_{21} + T_{12}a_{22} + T_{13}a_{23} > 0 \\ 2\pi - \arccos(T_{21}a_{12} + T_{22}a_{22} + T_{23}a_{32}) & \text{if } T_{11}a_{21} + T_{12}a_{22} + T_{13}a_{23} < 0 \end{cases} \quad (A.9)$$

All $\theta_C^{new1}, \theta_T^{new1}$ satisfying Eq. (A.9) above are solutions of Eq. (9).

For $\theta_G^{new1} = \pi$

$$\theta_C^{new1} - \theta_T^{new1} = \begin{cases} \arccos(T_{21}a_{12} + T_{22}a_{22} + T_{23}a_{32}) & \text{if } T_{11}a_{21} + T_{12}a_{22} + T_{13}a_{23} > 0 \\ 2\pi - \arccos(T_{21}a_{12} + T_{22}a_{22} + T_{23}a_{32}) & \text{if } T_{11}a_{21} + T_{12}a_{22} + T_{13}a_{23} < 0 \end{cases} \quad (A.10)$$

All $\theta_C^{new1}, \theta_T^{new1}$ satisfying Eq. (A.10) are solutions of Eq. (9)

The solutions for Eq. (14) are:

For the case $\sin\theta_G^{new} > 0$,

$$\theta_G^{new} = \arccos[T_{31}a_{31} + T_{32}a_{32} + T_{33}a_{33}] \quad (A.11),$$

$$\theta_C^{new} = \begin{cases} \arccos(-\dfrac{T_{11}a_{31} + T_{12}a_{32} + T_{13}a_{33}}{\sin\theta_G^{new}}), & \text{if } T_{21}a_{31} + T_{22}a_{32} + T_{23}a_{33} > 0 \\ 2\pi - \arccos(-\dfrac{T_{11}a_{31} + T_{12}a_{32} + T_{13}a_{33}}{\sin\theta_G^{new}}), & \text{if } T_{21}a_{31} + T_{22}a_{32} + T_{23}a_{33} < 0 \end{cases} \quad (A.12)$$

$$\theta_T^{new} = \begin{cases} \arccos(\dfrac{T_{31}a_{11} + T_{32}a_{12} + T_{33}a_{13}}{\sin\theta_G^{new}}), & \text{if } T_{31}a_{21} + T_{32}a_{22} + T_{33}a_{23} > 0 \\ 2\pi - \arccos(\dfrac{T_{31}a_{11} + T_{32}a_{12} + T_{33}a_{13}}{\sin\theta_G^{new}}), & \text{if } T_{31}a_{21} + T_{32}a_{22} + T_{33}a_{23} < 0 \end{cases} \quad (A.13)$$

We also have another set of solutions that are: $\theta_G^{new} \to 2\pi - \theta_G^{new}, \theta_C^{new} \to \theta_C^{new} + \pi, \theta_T^{new} \to \theta_T^{new}$

In other words, we always have two solutions for each set of rotation angles of a patient.

For the special case when $\cos\theta_G^{new} = \pm 1$, then $\theta_G^{new} = 0, \pi$.

For $\theta_G^{new} = 0$,

$$\theta_C^{new} + \theta_T^{new} = \begin{cases} \arccos(T_{21}a_{21} + T_{22}a_{22} + T_{23}a_{23}) & \text{if } T_{11}a_{21} + T_{12}a_{22} + T_{13}a_{23} > 0 \\ 2\pi - \arccos(T_{21}a_{21} + T_{22}a_{22} + T_{23}a_{23}) & \text{if } T_{11}a_{21} + T_{12}a_{22} + T_{13}a_{23} < 0 \end{cases} \quad (A.14)$$

All $\theta_C^{new}, \theta_T^{new}$ satisfying Eq. (A.14) above are solutions of Eq. (14).

For $\theta_G^{new} = \pi$

$$\theta_C^{new} - \theta_T^{new} = \begin{cases} \arccos(T_{21}a_{21} + T_{22}a_{22} + T_{23}a_{23}) & if \ T_{11}a_{21} + T_{12}a_{22} + T_{13}a_{23} > 0 \\ 2\pi - \arccos(T_{21}a_{21} + T_{22}a_{22} + T_{23}a_{23}) & if \ T_{11}a_{21} + T_{12}a_{22} + T_{13}a_{23} < 0 \end{cases} \quad (A.15)$$

All $\theta_C^{new}, \theta_T^{new}$ satisfying Eq. (A.15) are solutions of Eq. (14)

Table Caption: Tumor volume and position

| Patient No. | Tumor volume (cm$^3$) | Tumor Position |
|---|---|---|
| 1 | 0.848 | Right Frontal Lobe |
| 2 | 4.384 | Right Cerebellum |
| 3 | 2.515 | Right Frontal Lobe |
| 4 | 0.115 | Left Parietal Lobe |
| 5 | 5.359 | Cribriform plate |
| 6 | 2.360 | Parietal Lobe |
| 7 | 0.126 | Right Parietal Lobe |
| 8 | 2.171 | Left Occipital Lobe |

Figure Captions:

Figure 1: The ratio of the minimal dose of PTV for different rotation to its planning value is plotted as a function of rotation angles.

Figure 2: The ratio of the mean dose of PTV for different rotation to its planning value is plotted as a function of rotation angles.

Figure 3: The ratio of the maximal dose of PTV for different rotation to its planning value is plotted as a function of rotation angles.

Figure 4: The ratio of the conformity index of PTV for different rotation to its planning value is plotted as a function of rotation angles.

Figure 5: The ratio of the $V_{80\%}$ of PTV for different rotation to its planning value is plotted as a function of rotation angles.

Figure 6: The ratio of the $V_{95\%}$ of PTV for different rotation to its planning value is plotted as a function of rotation angles.

Figure 7: The ratio of the minimal dose of CTV for different rotation to its planning value is plotted as a function of rotation angles.

Figure 8: The ratio of the mean dose of CTV for different rotation to its planning value is plotted as a function of rotation angles.

Figure 9: The ratio of the maximal dose of CTV for different rotation to its planning value is plotted as a function of rotation angles.

Figure 10: The ratio of the $V_{95\%}$ of CTV for different rotation to its planning value is plotted as a function of rotation angles.

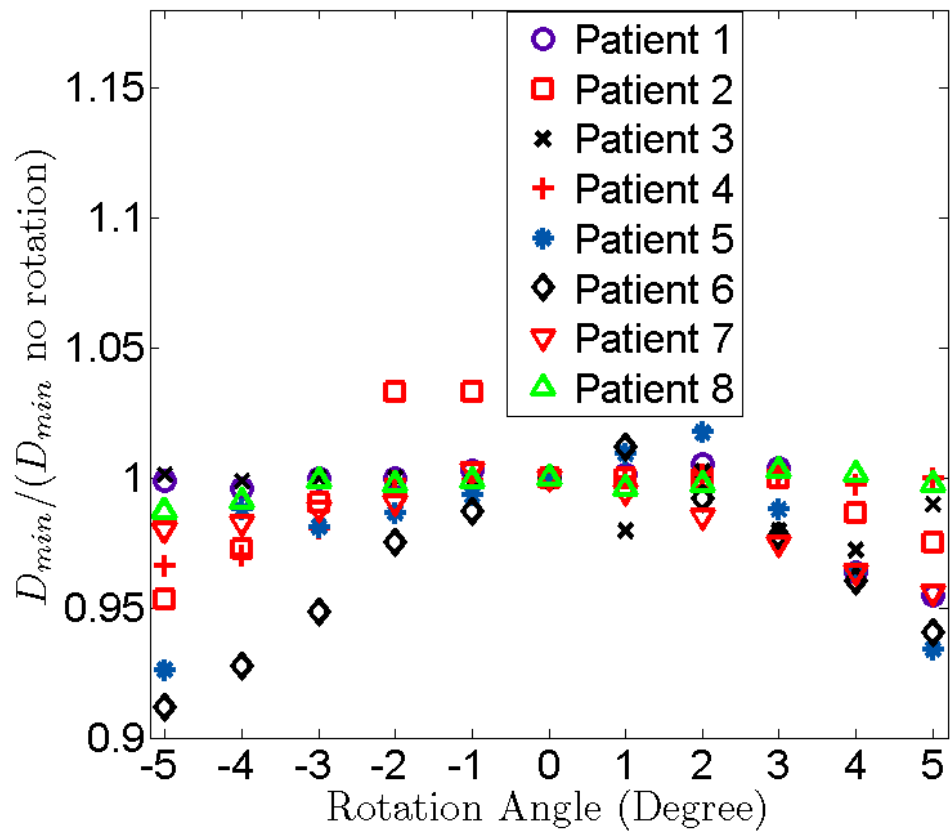

Fig. 1

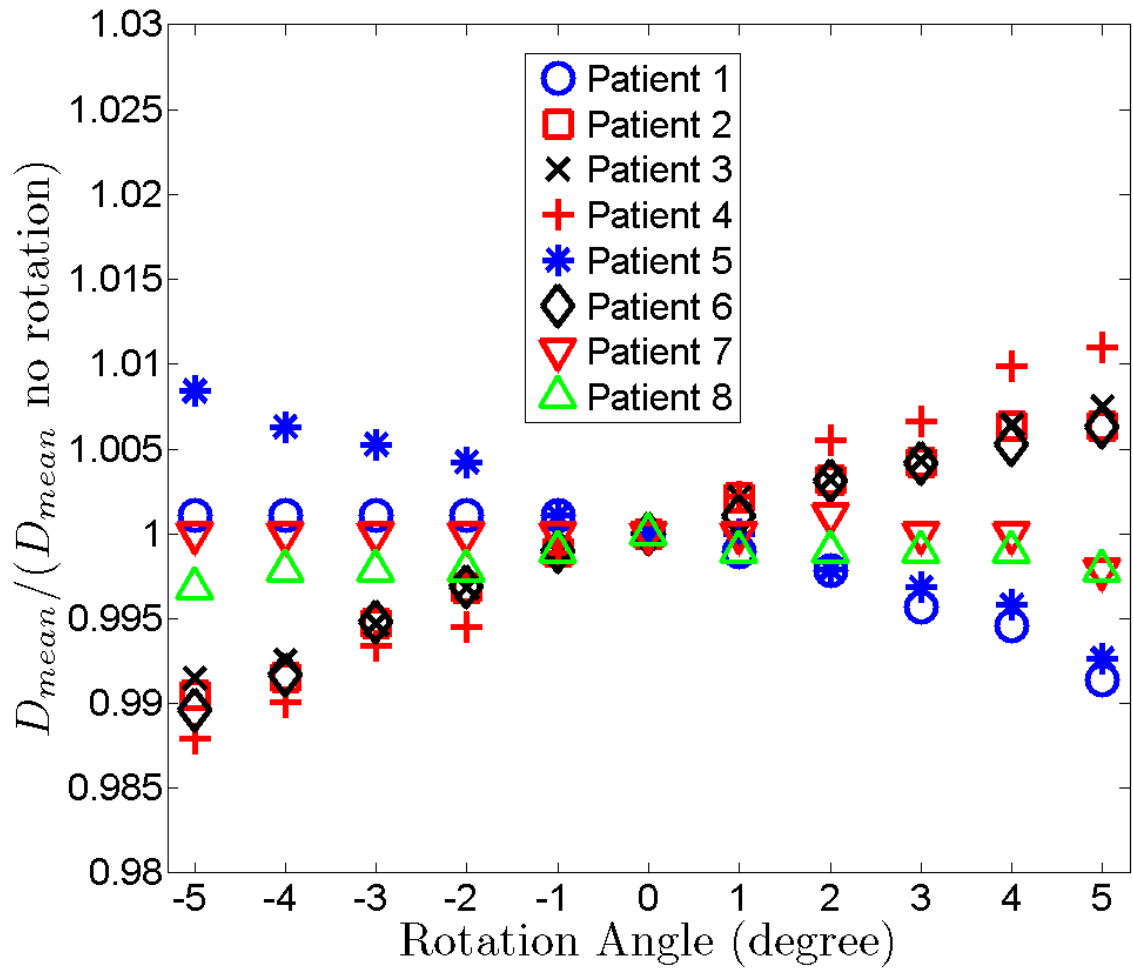

Fig.2

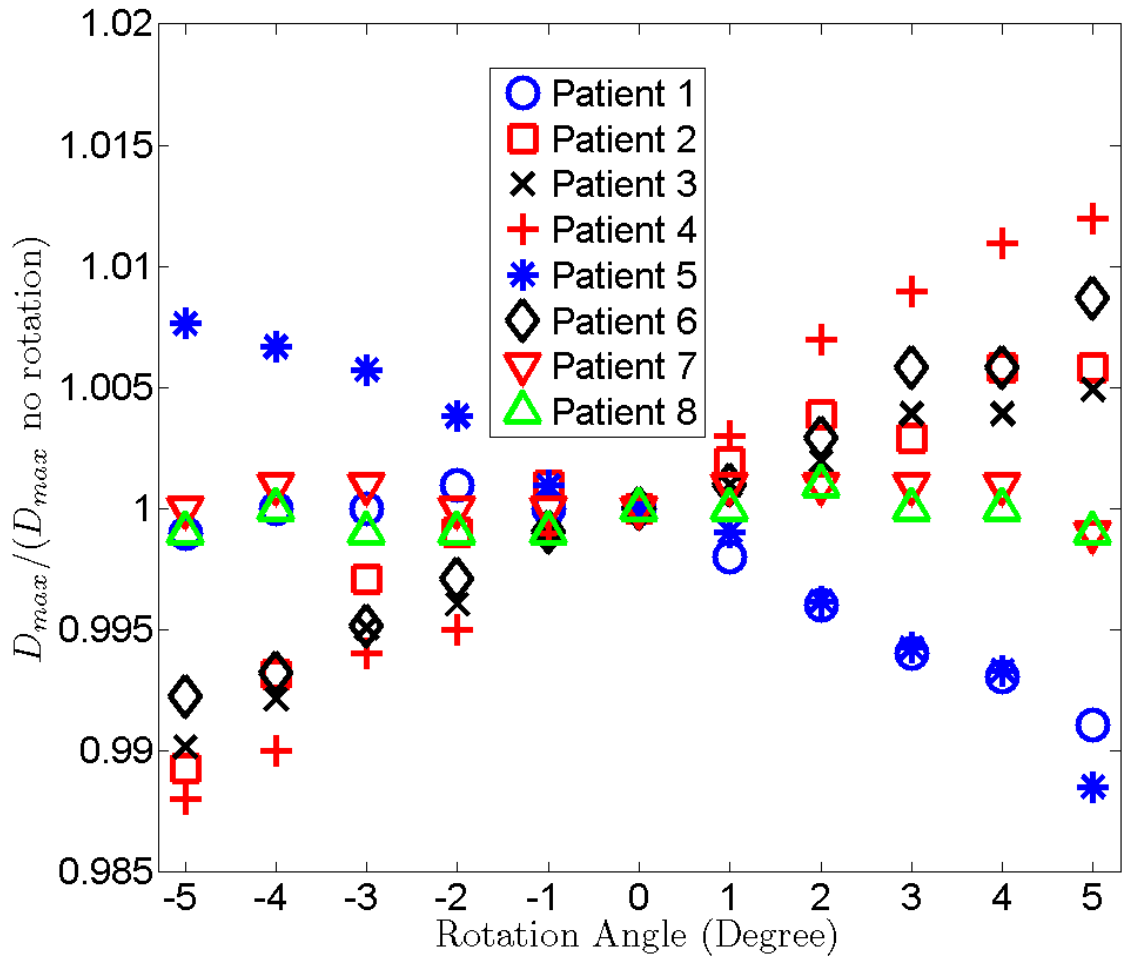

Fig. 3

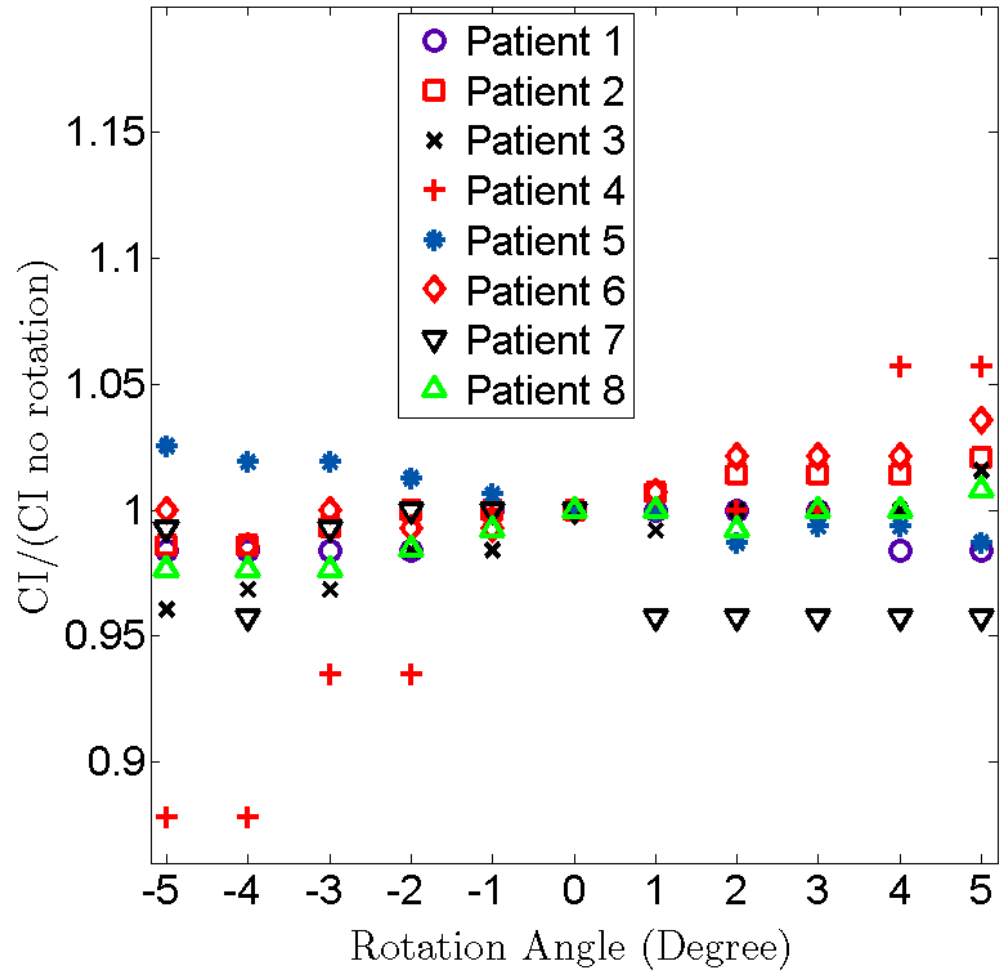

Fig. 4

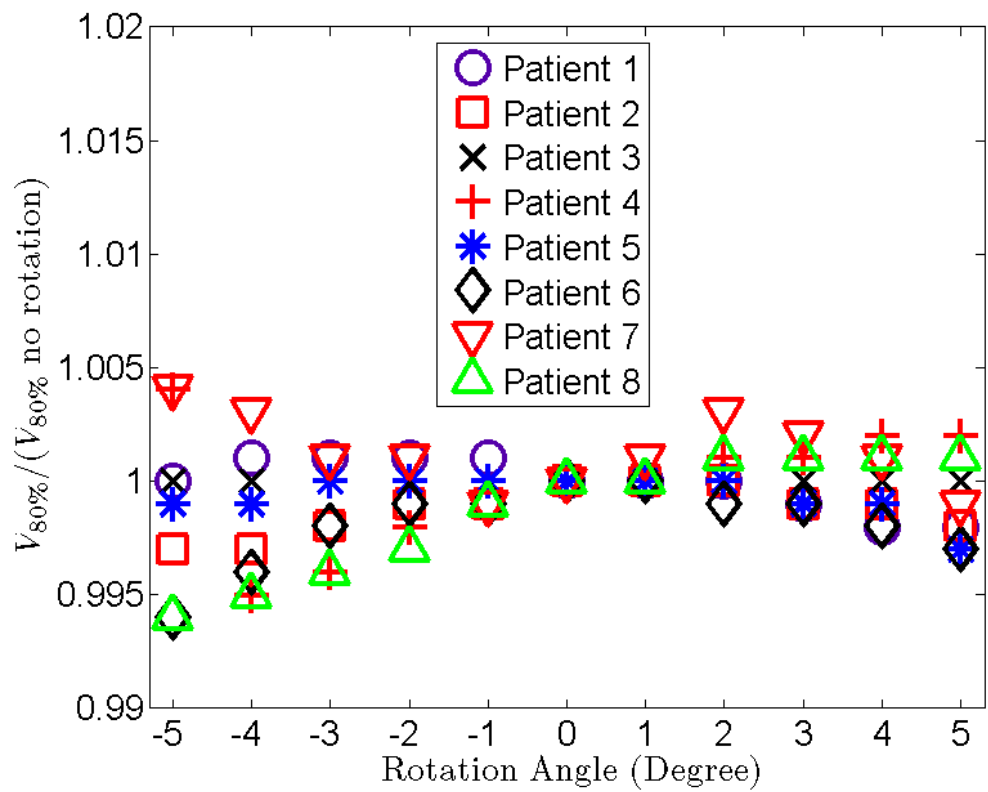

Fig. 5

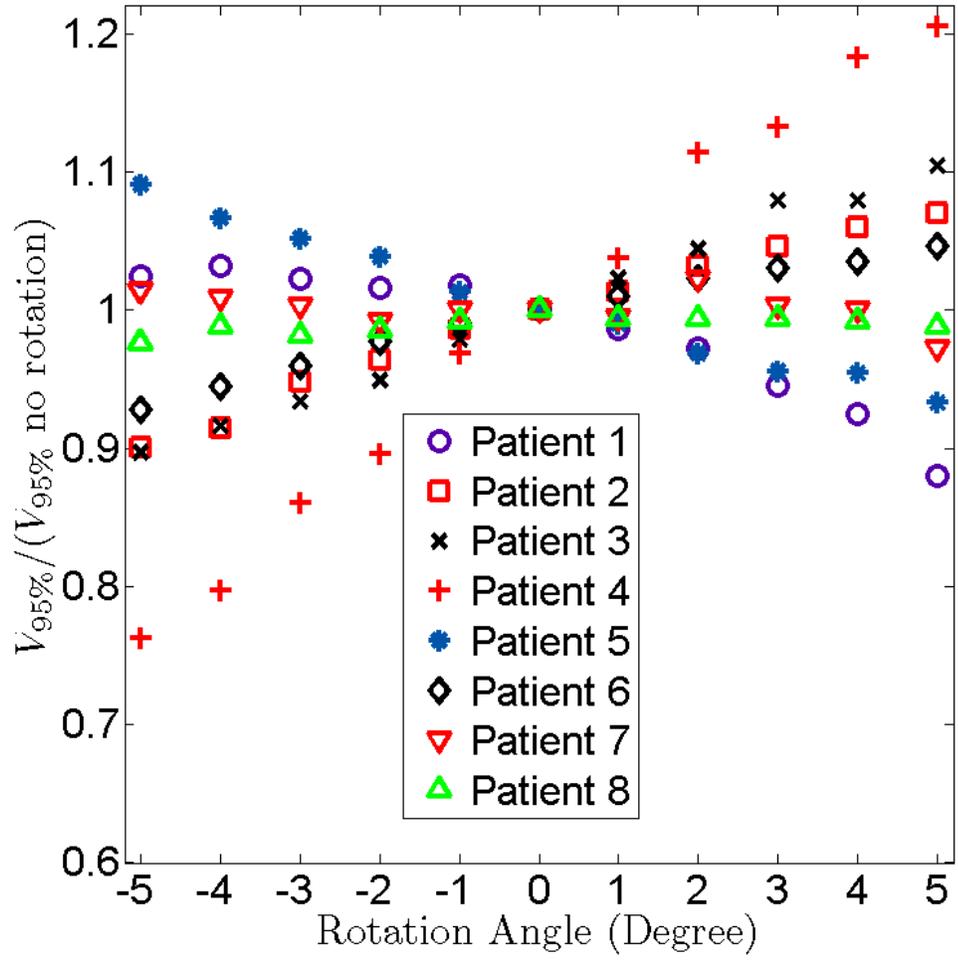

Fig. 6

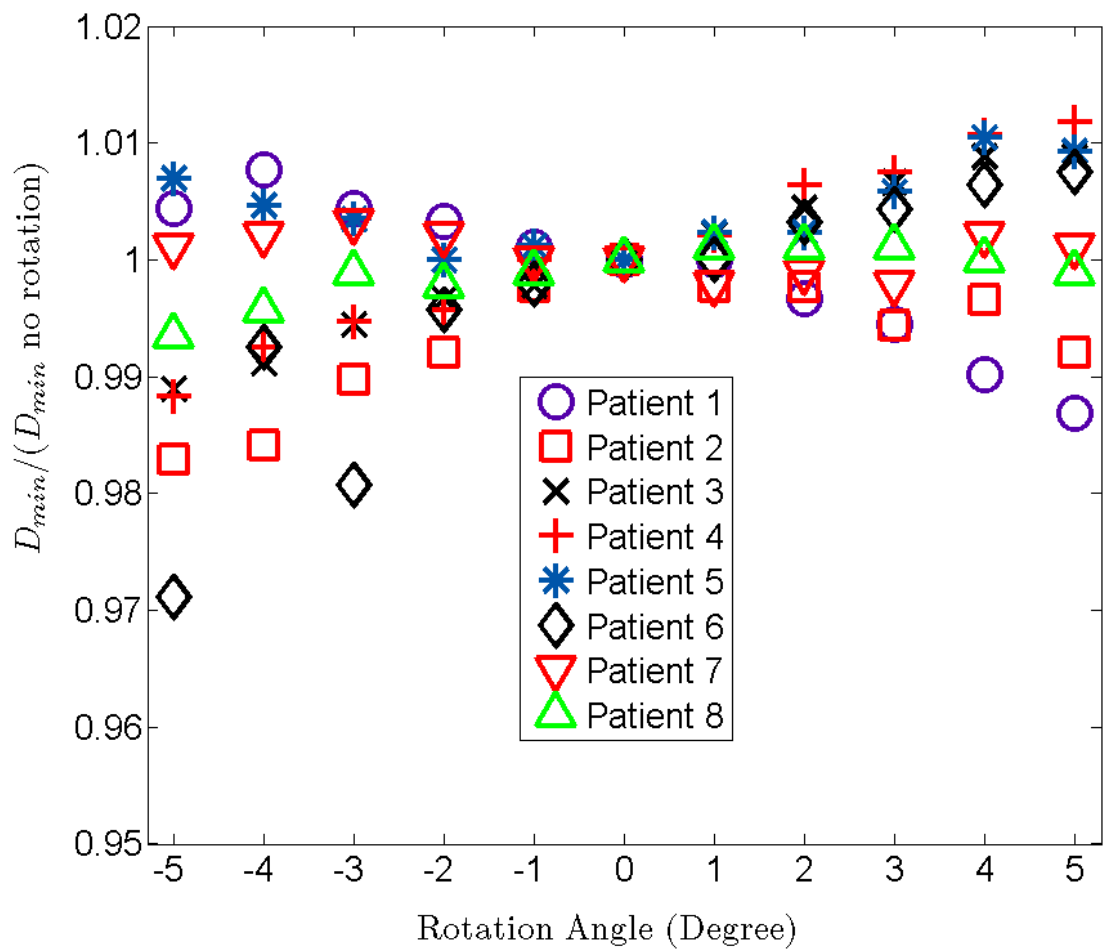

Fig. 7

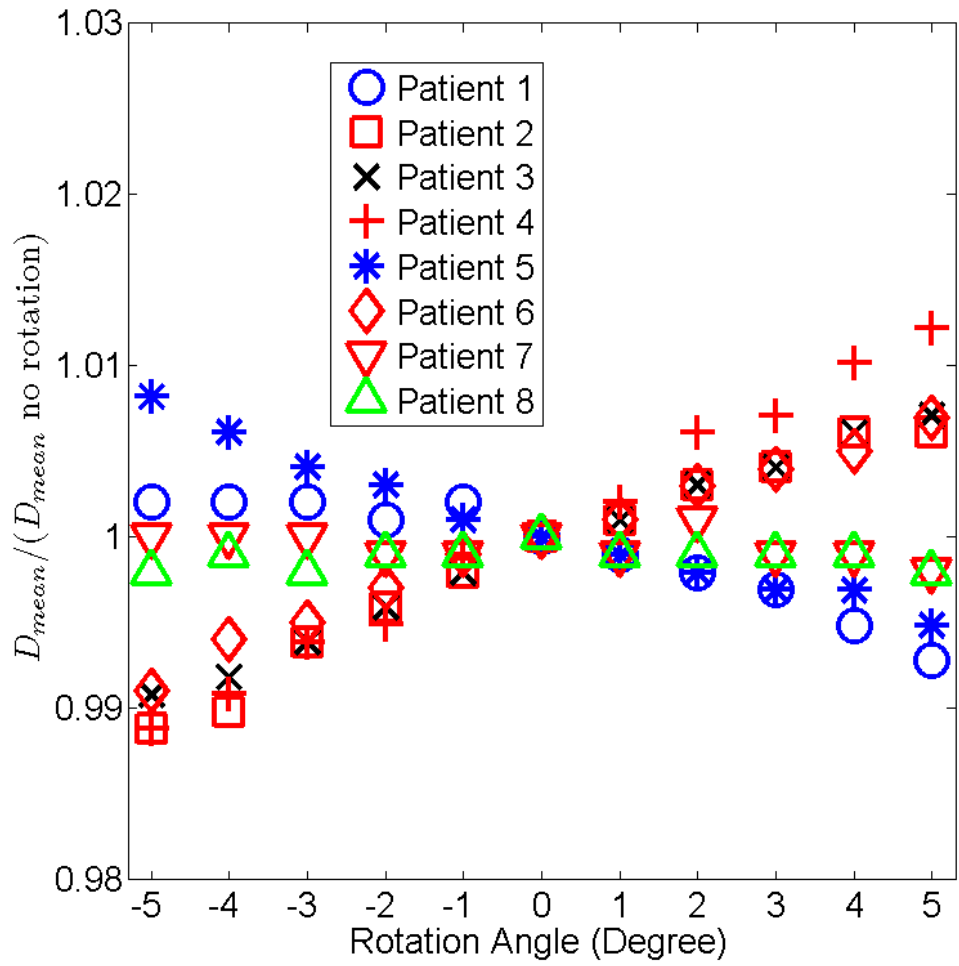

Fig.8

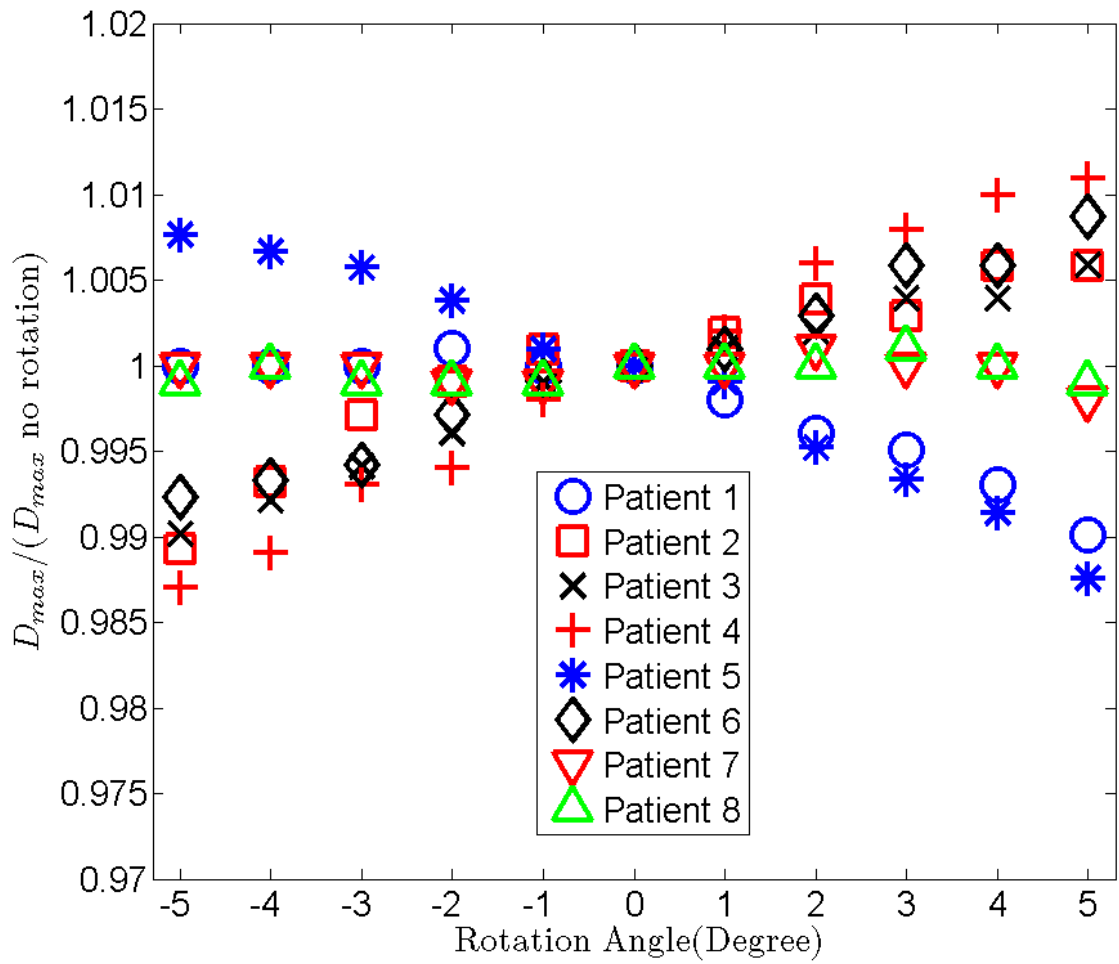

Fig.9

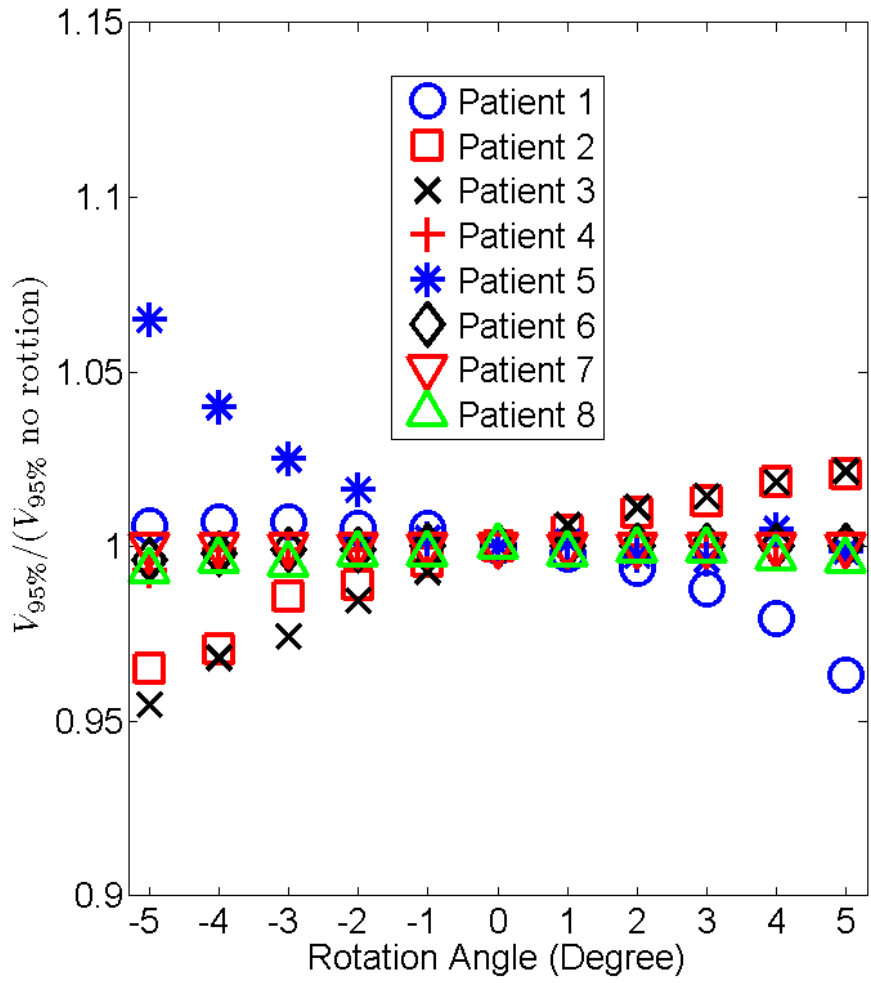

Fig.10